# Some Aspects of Intermediate mass black holes

C Sivaram and Kenath Arun[1]

Indian Institute of Astrophysics, Bangalore

**Abstract:** There is a lot of current astrophysical evidence and interest in intermediate mass black holes (IMBH), ranging from a few hundred to several thousand solar masses. The active galaxy M82 and the globular cluster G1 in M31, for example, are known to host such objects. Here we discuss several aspects of IMBH's such as their expected luminosity, spectral nature of radiation, associated jets, etc. We also discuss possible scenarios for their formation including the effects of dynamical friction, gravitational radiation, etc. We also consider their formation in the early universe and also discuss the possibility of supermassive black holes forming from mergers of several IMBH and compare the relevant time scales involved with other scenarios.

---

[1] Christ Junior College, Bangalore

Intermediate mass black holes (IMBH) are those black holes having mass between that of stellar black holes and supermassive black holes, i.e., in the range of 500 to $10^4 M_\odot$. Recent observations indicate an intermediate mass black hole in the elliptical galaxy NGC 4472, with an X-ray luminosity of $4 \times 10^{32}$ J/s.[1]

## 1. General considerations

The Schwarzschild radius corresponding to these masses are of the order of:

$$R_S = \frac{2GM}{c^2} \approx 1.5 \times 10^3 - 3 \times 10^4 \, km \qquad \ldots(1)$$

An accretion disc is formed by material falling into a gravitational source. Conservation of angular momentum requires that, as a large cloud of material collapses inward, any small rotation it may have will increase. Centrifugal force causes the rotating cloud to collapse into a disc, and tidal effects will tend to align this disc's rotation with the rotation of the gravitational source in the middle.

A black hole of mass M moving with a velocity $v$, in a medium of density, $\rho = nm_P$ the Bondi accretion rate is given by [2]: $\dot{m} = 4\pi R^2 n m_P V$ ...(2)

Where, $4\pi R^2$ is the cross section and it is given by: $4\pi R^2 = 4\pi \left(\frac{GM}{V^2}\right)^2$

The velocity is given by, $V^2 = c_S^2 + v^2$

Where, $c_S$ is the velocity of sound in the medium of density ρ and it is given by, $c_S = \sqrt{\gamma T R}$, and $\gamma = 5/3$ is the ratio of specific heats and $R$ is the universal gas constant.

The equation describing the velocity of isothermal winds has many solutions depending on the initial conditions at the base of the wind. There is only one critical solution for which the velocity increases from subsonic at the base to supersonic far out. This velocity passes through the critical point and implies one particular value of the initial velocity at the lower boundary of the isothermal region.

If the density at this point is fixed, the mass loss rate is fixed by equation (2) as, $\dot{m} = 4\pi R_0^2 n m_P V$. The total energy increases from negative at the base of the wind to positive in the supersonic region, so the flow requires the input of energy into the wind. This energy input is needed to keep the flow isothermal and it is this energy that is transferred into kinetic energy of the wind through the gas pressure.

As we shall see in the next section, for an intermediate mass black hole, the temperature is typically of the order of $10^4 K$. In this case, the velocity of sound $c_S = \sqrt{\gamma TR}$ is of the order of $10^4 m/s$.

Also we have, $R = \dfrac{2GM}{V^2}$, hence the accretion rate given by,

$$\dot{m} = \frac{4\pi n m_P (GM)^2}{V^3} \qquad \ldots(3)$$

For a typical number density of the order of $n = 10^{20} / m^3$, and the IMBH mass of $M = 10^4 M_\odot$ the accretion rate becomes, $\dot{m} = 10^{17} kg/s$.

The above expression implies that higher the density of the medium through which the black hole is travelling, more is the accretion rate. Also, the accretion rate is inversely related to the velocity with which the black hole is travelling in the medium. Although there is not much conclusive evidence for the IMBH, there is indirect evidence.

For a given mass of luminous object, there is a maximum value for its luminosity (Eddington luminosity), as the radiation pressure would tend to push the matter apart exceeding the gravitational force supporting it. This is the luminosity a body would have to have so that the force generated by radiation pressure exceeds the gravitational force. Thus, observed luminosity can set a lower limit on the mass of an accreting black hole.

*Chandra* and *XMM-Newton* observations in the nearby spiral galaxy have detected x-ray sources of luminosities of the order of $10^{33} W$, with the source away from the centre.[3] In section 4, we shall discuss this dynamics of the IMBH.

The X-ray luminosity corresponding to the accretion rate of $\dot{m} = 10^{17} kg/s$, is given by:
$L_X = \varepsilon \dot{m} c^2 \approx 5 \times 10^{31} W$. This matches with recent observed results.[4]

The Eddington luminosity is given by:

$$L_{Edd} = \frac{4\pi c G M m_P}{\sigma_T} \approx 10^{35} W \qquad \ldots(4)$$

For an IMBH of mass $10^4 M_\Theta$, where, the Thomson cross section is given by:

$$\sigma_T = \frac{8\pi}{3} \left( \frac{e^2}{m_e c^2} \right)^2 \approx 10^{-28} m^2$$

A possible candidate for the IMBH is the globular cluster G1 in M31, which is the most massive stellar cluster in the local group, with a mass of the order of $10^7 M_\Theta$.

Due to a large number of field stars contained within the accretion radius the Bondi accretion by an IMBH is complicated.[4] The accretion radius is given by:

$$R_{acc} = \frac{2GM_{BH}}{v^2} \approx 0.4 pc, \text{ for } M_{BH} = 10^4 M_\Theta; \quad v \approx 15 km/s$$

In the case of globular cluster G1 in M31, there are more than $10^5$ stars within 0.4pc of the centre.[5] The dynamical effects of these stars should also be considered.

Many more such dense stellar clusters are known to exist. For instance, a recently recognised super star cluster in our own galaxy is the Westerlund I. Most of its estimated half a million stars are crowded in a region hardly 3 parsecs wide. Several dozen of these stars are among the most massive and luminous superhot Wolf-Rayet stars, LBV, red supergiants, yellow hypergiants, etc. Also, near the Milky Way centre, we have the well-known Arches and Quintuplet clusters.

In the centre of M15 there are approximately five million stars per cubic parsec, which is hundred million times more stellar density than the solar neighbourhood. Also galaxies like M31, M33 and the Milky Way itself have comparable central stellar densities.

However M32 (satellite galaxy of Andromeda) has thirty million stars in a cubic parsec at its core. Even HST cannot resolve individual stars in this region.

More and more massive binary stars are being found, like for instance WR20a ($>80M_\Theta$ each, with 3.7 day period). As discussed above IMBH is likely to form in dense clusters containing young massive stars. IMBH is believed to power ultra luminous X-ray sources (ULXS). IMBH can capture companion stars, which provide accreting material to sustain ULX. These stars may be blue giants, white dwarfs, etc.

Tidal forces can rip giants and the material can fall into the IMBH. To rip apart these stars the mass of the black hole is around the mass of the IMBH.[6] The energy released during this process will be the binding energy of the stars which is given by:

$$E_{BE} = \frac{3}{5}\frac{GM^2}{R} \approx 10^{42} J$$

Several ULX is identified in the Antenna, a pair of colliding galaxies, producing several stars in dense cluster. Stellar collisions lead to formation of so-called megastars ($\sim 10^3 M_\Theta$), which collapses on short time scales to form IMBH. Clusters in M82, have such stellar densities. Binary and multiple IMBH can also form.

## 2. Black body considerations of IMBH

We have already obtained the Eddington luminosity as: $L_{Edd} = \frac{4\pi cGMm_P}{\sigma_T}$

By considering the black hole emission to obey black body radiation, we can use Stefan's law to relate the luminosity to the temperature, *T*, as:

$$\frac{4\pi cGm_P M}{\sigma_T} = \sigma A T_{max}^4 \qquad \ldots(5)$$

Where, σ is the Stefan's constant and area of the black hole and it is given by,
$$A = f(4\pi R_S^2) \qquad \ldots(6)$$

Where, $R_S = \dfrac{2GM}{c^2}$ is the Schwarzschild radius, $f$ indicates the size of the ambient gas around the black hole.

Making use of this in equation (5) we get,

$$T_{max}^4 = \left(\dfrac{c^5}{G}\dfrac{m_P}{f\sigma_T \sigma}\right) M \Rightarrow T_{max} \propto M^{-1/4} \qquad \ldots(7)$$

This implies that the temperature of the IMBH (accretion disk) of mass of the order of 500 to $10^4 M_\odot$ and $f$ ranging from 10 to 100 is:

$$T_{max} \approx 3 \times 10^6 K \qquad \ldots(8)$$

From Wien's law the corresponding wavelength is given by:

The corresponding wavelength is, $\lambda = \dfrac{(2.898 \times 10^{-3} Km)}{T} \approx 10^{-9} m \qquad \ldots(9)$

This lies in the soft X-ray region of the spectrum.[7]

During the earlier epochs of the universe, the density was much larger; hence the ambient density is also larger by the same factor of $(1+z)^3$.

The present density of the universe is of the order of one proton per cubic metre. As we shall see in section 5, the maximum redshift up to which we can detect supermassive black hole is of the order of $z = 12$.

The number density at this epoch is given by

$$n = 1 p/m^3 (1+z)^3 \approx 10^3 \, p/m^3 \qquad \ldots(10)$$

The temperature corresponding to this redshift and this number density is of the order of

$$T \approx 2 \times 10^7 K \qquad \ldots(11)$$

And the corresponding wavelength is of the order of $10^{-10} m \qquad \ldots(12)$

This falls in the X-ray region of the spectrum.

This wavelength is further red-shifted by a factor of $(1+z)$. Hence the observed wavelength will be of the order of $7 \times 10^{-8} m \qquad \ldots(13)$

This lies in the UV region of the spectrum.

## 3. Jets from the black hole

One of the manifestations of this accretion energy release is the production of so-called jets: the collimated beams of matter that are expelled from the innermost regions of accretion disks. These jets shine particularly brightly at radio frequencies. In rotating black holes, the matter forms a disk due to the mechanical forces present. In an Schwarzschild black hole, the matter would be drawn in equally from all directions, and thus would form an omni-directional accretion cloud rather than disk. Jets form in Kerr black holes (rotating black holes) that have an accretion disk.[8]

Black holes convert a specific fraction of accretion energy into radiation, which is traced by the X-ray luminosity and jet kinetic energy, which is traced by the radio-emission luminosity.[9] The matter is funnelled into a disk-shaped torus by the black hole's spin and magnetic fields, but in the very narrow regions over the black hole's poles, matter can be energized to extremely high temperatures and speeds, escaping the black hole in the form of high-speed jets. Inferred jet velocities close to the speed of light suggest that jets are formed within a few gravitational radii of the event horizon of the black hole.

The horizon for the Kerr black hole is given by,

$$r = m \pm \sqrt{m^2 - a^2} \qquad \ldots(14)$$

Here $m = \dfrac{GM}{c^2}$ is the geometric mass and $a = \dfrac{J_{MAX}}{Mc^2}$ is the geometric angular momentum.

From the condition that $r$ should be real, the limiting case is given by, $m = a$.
From this, the maximum angular momentum is given by,

$$J_{MAX} = \dfrac{M^2 G}{c} \qquad \ldots(15)$$

From the classical expression for the angular momentum associated with a jet of length $l$, assuming the particles to be travelling at near speed of light, the expression becomes

$$J = mcl \qquad \ldots(16)$$

Considering a conical jet with base radius $r$ and density $\rho$, the mass of the jet is given by,

$$m = \frac{1}{3}\pi r^2 l \rho \qquad \ldots(17)$$

Then the angular momentum becomes

$$J = \frac{1}{3}\pi l^2 r^2 c \rho \qquad \ldots(18)$$

From the geometry of the jet, we can relate the length of the jet to the radius $r$ as $r = l \tan 5^0$. Here we have assumed the small opening angle of the jet to be $5^0$.

Length of the jet, $l = \left( \frac{3GM^2}{\pi \rho c^2 (\tan 5)^2} \right)^{1/4} \qquad \ldots(19)$

For typical densities of the ambient gas and for the IMBH of mass $10^4 M_\Theta$, the length is of the order of 20pc.

## 4. Evolution of a star cluster and possible scenario for IMBH formation

One of the possible models for the formation of an IMBH is the collapse of a cluster of stars.[10] The collapsed core can accrete matter ejected during the formation. If a collection of a thousand 10 solar mass stars in a volume of a parsec cube collapses, ejecting 30% of its mass, then the total mass of the ambient gas is given by:

$$nm_P = 0.3 \times 10^3 \times 10 M_\Theta \approx 6 \times 10^{33} kg \qquad \ldots(20)$$

The accretion rate is given by equation (3) as: $\dot{m} = \frac{16\pi n m_P G^2 M^2}{c^3}$

If 30% of the mass of each star is ejected, then this implies an accretion rate of $\dot{m} \approx 4 \times 10^7 kg/s$. And the corresponding (Eddington) luminosity is:

$$L_{Edd} = \frac{GM\dot{m}}{\sigma_T R_S^2 n} \approx 10^{24} W \qquad \ldots(21)$$

Dynamical friction is related to loss of momentum and kinetic energy of moving bodies through a gravitational interaction with surrounding matter in space. The effect must exist if the principle of conservation of energy and momentum is valid since any gravitational interaction between two or more bodies corresponds to elastic collisions between those bodies. For example, when a heavy body moves through a cloud of lighter bodies, the gravitational interaction between this body and the lighter bodies causes the lighter bodies to accelerate and gain momentum and kinetic energy.

Since energy and momentum are conserved, this body has to lose a part of its momentum and energy equal to the sums of all momenta and energies gained by the light bodies. Because of the loss of momentum and kinetic energy of the body under consideration the effect is called dynamical friction. Of course the mechanism works the same way for all masses of interacting bodies and for any relative velocities between them.

However, while in the above case the most probable outcome is the loss of momentum and energy by the body under consideration, in the general case it might be either loss or gain. In a case when the body under consideration is gaining momentum and energy the same physical mechanism is called sling effect.

The full Chandrasekhar dynamical friction formula for the change in velocity of the object involves integrating over the phase space density of the field of matter.[11] By assuming a constant density though, a simplified equation for the force from dynamical friction, $f_d$, is given as:

$$f_d \approx C \frac{(GM)^2 \rho}{v_M^2} \quad \ldots(22)$$

Where, $G$ is the gravitational constant, $M$ is the mass of the moving object, $\rho$ is the density, and $v_M$ is the velocity of the object in the frame in which the surrounding matter was initially at rest. In this equation $C$ is not a constant but depends on how $v_M$ compares to the velocity dispersion of the surrounding matter. The greater the density of the surrounding media, stronger will be the force from dynamical friction. Similarly, the force is proportional to the square of the mass of the object. The force is also proportional to the inverse square of the velocity. This means the fractional rate of energy loss drops rapidly at high velocities.

Dynamical friction is, therefore, unimportant for objects that move relativistically, such as photons. Dynamical friction is particularly important in the formation of planetary systems and interactions between galaxies.

During the formation of planetary systems, dynamical friction between the protoplanet and the protoplanetary disk causes energy to be transferred from the protoplanet to the disk. This results in the inward migration of the protoplanet.

When galaxies interact through collisions, dynamical friction between stars causes matter to sink toward the centre of the galaxy and for the orbits of stars to be randomised. The dynamical friction comes into effect in the evolution of cluster between the ambient gas and dust and the central IMBH.

If the IMBH of mass $M_{BH}$ is moving with a velocity of $v_b$ in a uniform background of 'fixed' lighter stars of equal masses *m*. As $M_{BH}$ moves, a star approaching with impact parameter *b*, will have a velocity change given by: $\Delta v \approx a \Delta t$.

Where, $\Delta t$ is the encounter duration, $a$ is the acceleration. They are given by:

$$ma = \frac{GmM_{BH}}{b^2}; \quad \Delta t = \frac{b}{v_b} \qquad \ldots(23)$$

The change in velocity becomes: $\Delta v \approx a\Delta t \approx \frac{GM_{BH}}{b^2}\frac{b}{v_b}$ ...(24)

The kinetic energy gained by the star corresponding to this change in velocity is given by:

$$\Delta E = \frac{1}{2}m(\Delta v)^2 \approx \frac{1}{2}m\left(\frac{GM_{BH}}{bv_b}\right)^2 \qquad \ldots(25)$$

If *n* is the number density of the stars, then the number of encounters with impact parameter between $b + \Delta b$ and *b* is given by:

$$\Delta N \approx n(v_{BH}\Delta t)\Delta(\pi b^2) \qquad \ldots(26)$$

The total change in velocity is given by:

$$\frac{dv_{BH}}{dt} \approx \frac{1}{M_{BH} v_{BH}} \int \frac{dE}{dt} dN \approx \frac{\pi G^2 n M_{BH}}{v_{BH}^2} \int_{b_1}^{b_2} \frac{db}{b} \quad \ldots(27)$$

Where, $b_1$, the lower bound on $b$, is given for the case where the gravitational energy is of the order of the kinetic energy of the black hole. That is,

$$\frac{G m M_{BH}}{b_1} \approx \frac{1}{2} m v_{BH}^2 \quad \ldots(28)$$

And the upper limit $b_2$ is the size of the system. Let $\int_{b_1}^{b_2} \frac{db}{b} = \ln \Lambda$

Then the total change in velocity of the black hole is given by:

$$\frac{dv_{BH}}{dt} \approx \frac{\pi G^2 n M_{BH}}{v_{BH}^2} \ln \Lambda \quad \ldots(29)$$

In the above discussion we have assumed that the stars are stationary. This need not be true. If the stars have a velocity dispersion of $\sigma$ and the black hole is moving at a slower velocity than this, i.e., $v_{BH} << \sigma$, then the dynamical friction force is given by:

$$F_{DF} = m a_{DF} \approx \frac{-3\pi G^2 n m M_{BH}^2 \ln \Lambda}{\left(\sqrt{2}\sigma\right)^3} v_{BH} = -\gamma v_{BH} \quad \ldots(30)$$

Where, $\gamma = \frac{-3\pi G^2 n m M_{BH}^2 \ln \Lambda}{\left(\sqrt{2}\sigma\right)^3} \quad \ldots(31)$

In the case of the velocity of the black hole being faster than $\sigma$, the dynamical friction will be reduced and the black hole will slide through the cluster.

The black hole is also subjected to a gravitational force due to the star cluster, which is given by,

$$\nabla^2 \phi(r) = 4\pi G \rho(r) = 4\pi G m n(r) \quad \ldots(32)$$

Where, $n(r)$ is the density distribution of stars and m is the typical mass of the star (assuming all stars are of same mass).

For a constant density, $\rho(r) = \rho_0$, the potential is given by:

$$\phi(r) = -2\pi G \rho_0 \left( R^2 - \frac{1}{3} r^2 \right) \quad \ldots(33)$$

Where, $R$ is the radius of the cluster.

The gravitational force on the black hole is given by:

$$F_g = -M_{BH} \nabla \phi(r) = -\frac{4}{3} \pi G \rho_0 M_{BH} r = -kr \quad \ldots(34)$$

(The particle inside a homogenous gravitational system performs simple harmonic motion!)

The equation of motion of the black hole in the star cluster is given by:

$$M_{BH} \frac{d^2 r}{dt^2} + kr + \gamma \frac{dr}{dt} = 0 \quad \ldots(35)$$

This corresponds to a damped oscillator. And the solution is given by: [12]

$$r = M_{BH} \exp\left(-\gamma t / 2M_{BH}\right) \cos\left(\left(\sqrt{\frac{k}{M_{BH}}}\right) t + \gamma\right) \quad \ldots(36)$$

The black hole undergoes a damped oscillation in the star cluster. The damping time corresponding to the system is given by: $t = M_{BH}/\gamma$, where, $\gamma = \dfrac{-3\pi G^2 nm M_{BH}^2 \ln \Lambda}{\left(\sqrt{2}\sigma\right)^3}$.

In the case of the system M82, which harbours an IMBH of mass in the range of 500 to $10^4 M_\odot$, the black hole is not found at the centre, but displaced by about one kilo-parsec from the centre.

The period corresponding to the oscillation is given by:

$$T = \frac{2\pi}{\omega}; \quad \omega = \sqrt{\frac{k}{M_{BH}}} \approx 10^{-13} s^{-1} \quad \ldots(37)$$

From equation (32), we can work out the time taken for the BH to shift by this distance. Using this equation along with (27) and (30), we get the time of the order of $10^6$ years.

For the system under consideration, the number density of the stars in the cluster is $n \approx 10^4/(pc)^3$, typical mass of the star in the cluster is about one solar mass, and the velocity dispersion is of the order of $\sigma \approx 30 km/s$.

We get the damping time for the system as:

$$t = \frac{\left(\sqrt{2}\sigma\right)^3}{3\pi G^2 nmM_{BH} \ln \Lambda} \approx 4 \times 10^{12} s \qquad \ldots(38)$$

This works out to be of the order of $10^5$ years.

Taking the effects of dynamical friction into consideration, the relaxation time for the system to form the IMBH is given by:

$$t = \frac{v^3}{nG^2 M_{BH}^2 \ln N} \approx 10^7 \text{ years} \qquad \ldots(39)$$

For a denser core, i.e., about $10^3 \text{ stars}/(0.01 pc)^3$, the time taken to form the IMBH is given by,

$$t = \frac{v^3}{nG^2 M_{BH}^2 \ln N} \approx 4 \times 10^{12} s = 10^6 \text{ years} \qquad \ldots(40)$$

### 5. Formation of supermassive black hole (SBH): Merger of IMBH

Portegies Zwart, et al[13] propound that dense star densities near galactic centres can lead to runaway stellar mergers thus efficiently producing IMBH's. Indeed they suggest that the inner most ten parsecs of our galaxy can contain about 50 IMBH's, each of about thousand solar masses. These can sink towards the core as they interact with the stars. Every time an IMBH has a stellar encounter, the stars' velocity is boosted. This reduces the potential energy making the IMBH sink to the galactic core. These IMBH's can thus ultimately merge with the galaxy's supermassive black hole.

However if very near the galaxy SBH there are not enough stars, the IMBH may stop falling inward, halting at about 0.01 light years from the centre. However if there are several IMBH's near the galactic core, they can interact with one another and once every million years an IMBH can merge with the SBH.

It has been suggested that presence of several IMBH's near the galactic centre might explain why there are so many clusters of young stars (like S2, the B type star) in the galactic core, where tidal forces should rip apart the gas clouds from which the stars form.

These stars could have formed much further away and the IMBH's could have shepherded them inwards, much like the thin rings of Uranus or Saturn that are kept in place by being 'shepherded' by tiny satellites.

This IMBH can merge with the stars in the surrounding volume, $\left(\sim 10^6 M_\odot/(1pc)^3\right)$, to give a supermassive black hole in the time scale given by,

$$t = \frac{v^3}{nG^2 M^2 \ln N} \approx 10^{15} s = 10^8 \, years \qquad \ldots(41)$$

The maximum redshift observed till now is of the order of $z = 6.3$. The age of the universe corresponding to this redshift is given by:

$$t = \frac{1}{H_0}\left(\frac{1-(1+z)^{-1/2}}{1+z}\right) \approx 10^9 \, yrs \qquad \ldots(42)$$

According to the model suggested above, the time taken for the formation of SBH is of the order of $10^8$ years. The corresponding redshift is of the order of $z = 12$. For redshifts above this limit, as per this model, we should not be able to detect any supermassive black holes.

Other possible ways in which SBH can form is: (1) By the merger of two or more IMBH. Moving masses like black holes produce gravitational waves in the fabric of space-time. A more massive moving object will produce more powerful waves, and objects that move very quickly will produce more waves over a certain time period. (2) By accretion of matter by the IMBH in systems such as AGNs and quasars.

Gravitational waves are usually produced in an interaction between two or more compact masses. Such interactions include the binary orbit of two black holes orbiting each other. As the black holes orbit each other, they send out waves of gravitational radiation that reaches the Earth, however, once the waves do get to the Earth, they are extremely weak.

This is because gravitational waves decrease in strength as they move away from the source. Even though they are weak, the waves can travel unobstructed within the fabric of space-time.

From Kepler's third law, the period is related to the separation by,

$$P^2 = \frac{4\pi^2}{G(M_1 + M_2)} R^3 \qquad \ldots(43)$$

Knowing the period, we can determine the orbital velocity from, $vP = 2\pi R$

The energy lost by the system due to the emission of gravitational waves, for a circular orbit, is given by,

$$\dot{E}_{GW} = \frac{128 v^{10}}{5 G c^5} \qquad \ldots(44)$$

For two IMBHs of mass $M = 10^4 M_\odot$ each, to merge in $\tau_{mer} \approx 10^9$ years, the distance of separation should be of the order of, 

$$R = \frac{GM^2}{\tau_{mer} \dot{E}_{GW}} \qquad \ldots(45)$$

Where, $\dot{E}_{GW} = \frac{128 v^{10}}{5 G c^5}$, $v = \frac{2\pi R}{P}$, $P = \left( \frac{4\pi^2}{G(M_1 + M_2)} R^3 \right)^{1/2} \Rightarrow v = \sqrt{\frac{2GM}{R}} \qquad \ldots(46)$

And the energy loss due to the gravitational waves emission is given by,

$$\dot{E}_{GW} = \frac{128}{5Gc^5}\left(\frac{2GM}{R}\right)^5 \quad \ldots(47)$$

Therefore the distance of separation is given by:

$$R \approx \left(8\times10^2 \frac{(GM)^3 \tau_{mer}}{c^5}\right)^{1/4} \approx 2\times10^{12}\, m \quad \ldots(48)$$

The corresponding orbital frequency is given by, $f = \frac{v}{2\pi R}$ ...(49)

The orbital velocity is of the order of, $v = \sqrt{\frac{2GM}{R}} \approx 10^6\, m/s$ ...(50)

This implies that the frequency is, $f \approx 10^{-7}\, Hz$, and hence the period will be given by, $P = 10^7\, s = 1\, year$.

For two IMBH with separation of about $10^{-4}$ parsec, the time taken to merge is of the order of Hubble time. For such a model to produce a SBH, we need about $10^3$ IMBH merging together.

Hence the model discussed earlier, with the IMBH merging with the surrounding stars, gives a much more efficient way of generating a supermassive black hole.

In the case of accretion of matter by IMBH to form SBH, the increase in mass is exponential with time:

$$M = M_0 \exp(kt) \quad \ldots(51)$$

Where, $k^{-1} = t_0 \approx 6\times10^8$ years is the characteristic time required for the mass to increase $e$ folds, with the accreting disk emitting at maximum luminosity.

For the IMBH to accrete enough matter to become SBH of mass say $10^8 M_\odot$, we have:

$$\exp(kt) = \frac{M}{M_0} = 10^4 \quad \ldots(52)$$

The corresponding time scale is of the order of $t \approx 5 \times 10^9$ years ...(53)

This implies that the mass will have to increase $\approx e^{10}$ folds by accretion for the IMBH to become an SBH. Hence even this model does not provide an efficient way of formation of SBH from an IMBH.

IMBH's would have formed in the early universe. Owing to low metal content, the earliest stars would have been very massive, a few hundred solar masses.[14] Such stars would end up in a pair-instability supernova (around oxygen-neon burning temperature of 2 billion degrees) and would collapse into a black hole (if their mass exceeds 250 solar mass).

Signatures of such explosions of supermassive stars at $z \approx 10$ (when the universe was only half a billion years old, and eleven times smaller) could be sought with future space telescopes.[15, 16] (There are reports of galaxies at $z \approx 10$, with the VLT)

Observations of galaxy core show correlation between black hole masses and the spheroidal (bulge) component of the host galaxy. Thus the primordial low mass galaxies (blue galaxies) would host low mass central black holes, which just correspond to the IMBH mass (about $10^{-5}$ to $10^{-6}$ the galaxy mass).

Merger of these primeval galaxies would lead to larger galaxies and cluster and the IMBH's would also merge and sink to the core forming an SBH (like for instance M87 has a $3 \times 10^9 M_\odot$ black hole).

## 6. Concluding remarks

We have discussed several aspects of the expected characteristics of IMBH like their luminosity, accretion rate, formation, etc. They are likely to have formed in the early universe, around $z = 12$. Supermassive black holes are not likely to form from merger of IMBH's. Other possible scenarios are also discussed.